\documentclass[12pt]{iopart}
\usepackage{iopams}  
\usepackage{graphicx}

\begin{document}

\title[Massive QNM]{Massive quasi-normal mode}

\author{Akira Ohashi\dag\ and Masa-aki Sakagami\ddag}

\address{\dag\ Department of Physics, Ochanomizu University,\\
2-1-1, Otsuka, Bunkyo, Tokyo, 112-8610, Japan}

\address{\ddag\ Graduate School of Human and Environmental Studies,
Kyoto University,\\
Yoshida-Nihonmatsu, Sakyo, Kyoto 606-8501, Japan}

\begin{abstract}
This paper purposes to study quasi-normal modes due to massive scalar fields.
We, in particular, investigate the dependence of QNM frequencies on
the field mass.
By this research, we find that
there are quasi-normal modes with arbitrarily long life
when the field mass has special values.
It is also found that QNM can disappear
when the field mass exceed these values.
\end{abstract}




\section{Introduction}
Quasi-normal mode (QNM) is one of the important and exciting themes
in the black hole physics.
QNMs represent the behaviours of fields on a black hole spacetime
under a certain condition, imposed both at the event horizon of
the black hole and at the asymptotic infinity of the spacetime.
This means that QNMs can be strongly affected
by the curvature of the spacetime.
In other words, we can expect that
the property of QNMs strongly reflects
the property of the spacetime, 
and so we can grasp the property of the spacetime
by studying on QNMs.

The existence of QNMs was mentioned first by Vishveshwara\cite{KokkotasSchmidt}
in the context of scattering problem by black holes.
Later, people began to perceive that QNMs were important sources
of gravitational waves, and many researchers developed various 
methods to calculate QNM frequencies \cite{Chandra}, \cite{Leaver},
\cite{WKB}, \cite{WKB2}, \cite{BlomeMashhoon}, \cite{MajPan},
\cite{FerrariMashhoon}, \cite{AndAraSch}.
Recently, some researchers have pointed out that
the asymptotic behaviour of the distribution of QNMs
relates to the quantum gravitational theory
\cite{Hod1}, \cite{Nollert}.
Thus, we recognise that researches on properties of QNMs will
further increase its importance.

Our motivation of this article is to study properties of a black hole spacetime
using QNMs as a mathematical tool.
Almost all studies on QNMs discuss only massless fields such as
electromagnetic waves and gravitational waves,
since these studies mainly focuses on phenomenological significances.
Therefore, there are few studies on QNMs by massive fields
(hereafter, we call them massive QNMs).
One of the studies related to massive QNMs is
a work by Simone and Will \cite{SimoneWill},
which investigates massive QNMs on black hole spacetime using WKB method.
They study the dependency of QNM frequencies on the mass of the field,
but their discussion is restricted to narrow range of the field mass
due to the restriction required by the WKB method.
Hence, their analysis cannot fully reveal
the dependency of QNM frequencies on the field mass.
Another work on massive QNM is a study by Konoplya \cite{Konoplya}.
He study massive QNMs on a charged black hole back ground and
consider QNMs due to massive and charged fields.
His paper mainly focuses on correlation between QNMs and
the charge of the field and the black hole.
Therefore, the dependency of QNMs on the field mass
is not argued , though his article investigate 
almost same situation that we will study in this article.

Hence, we will study behaviours of QNMs to the field mass in detail.
For this purpose, we first consider a model in which we can derive
QNM frequencies analytically. Through this investigation we find that
there is a singular phenomenon that QNMs may disappear
when the field mass becomes sufficiently large.
Second, we study QNMs on the Reisner-Nordstr\"om black hole spacetime
by a numerical method.
As a result, we confirmed that the singular phenomenon
observed in the model can also be observed in the real black hole case.

This article consists of the following contents.
In \sref{sec:am},
we study a correlation between
QNM frequencies and the field mass
in a model which can be solved analytically.
In \sref{sec:bh},
we examine the real black hole case with a numerical method.
In \sref{sec:ds}, we discuss the result obtained in \sref{sec:bh}.
In \sref{sec:sum}, we summarise our investigation.
Additionally, \ref{sec:Leaver} shows the review of
the calculation algorithm used in \sref{sec:bh}.

\section{Analytical model}\label{sec:am}
In this section, we study the dependence of massive QNM frequencies
on the field mass by using an analytically solvable model.

Generally, in order for QNMs to exist, it is necessary that
the background has a potential that can trap waves.
Then, the trapped waves will exude out gradually.
This phenomenon is exactly QNMs.
If the asymptotic values of the potential are the same
at both side of the potential peak,
the waves will be the massless QNMs.
If the asymptotic values are different,
they will be massive QNMs.

\begin{figure}
\begin{center}
\includegraphics[height=60mm]{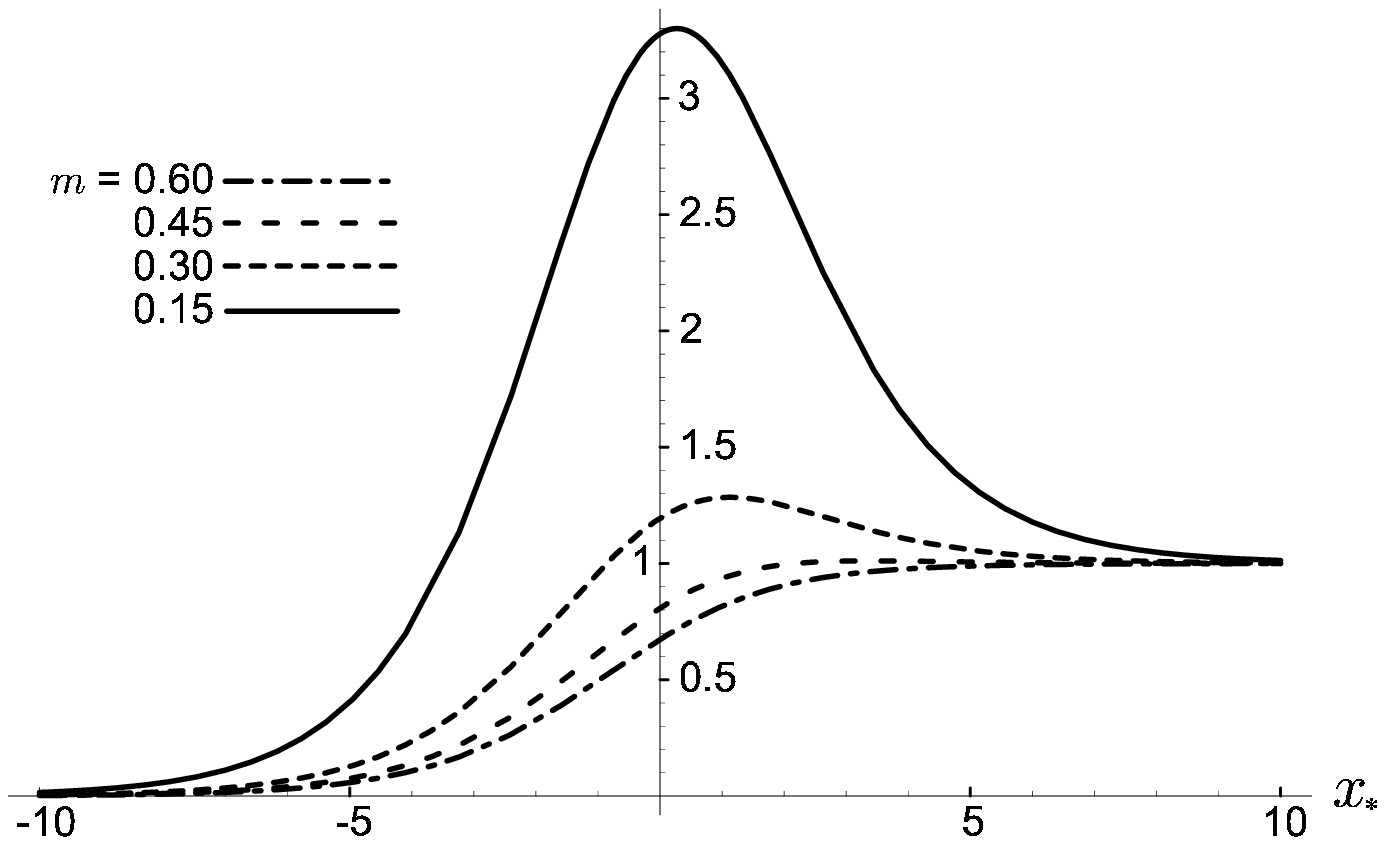}
\caption{\label{caption:poth}
Profile of the potential (\ref{eq:hppot}) in the analytic model for $M=1$.
The potential is normalised by $m^2$.}
\vspace{4mm}
\includegraphics[height=60mm]{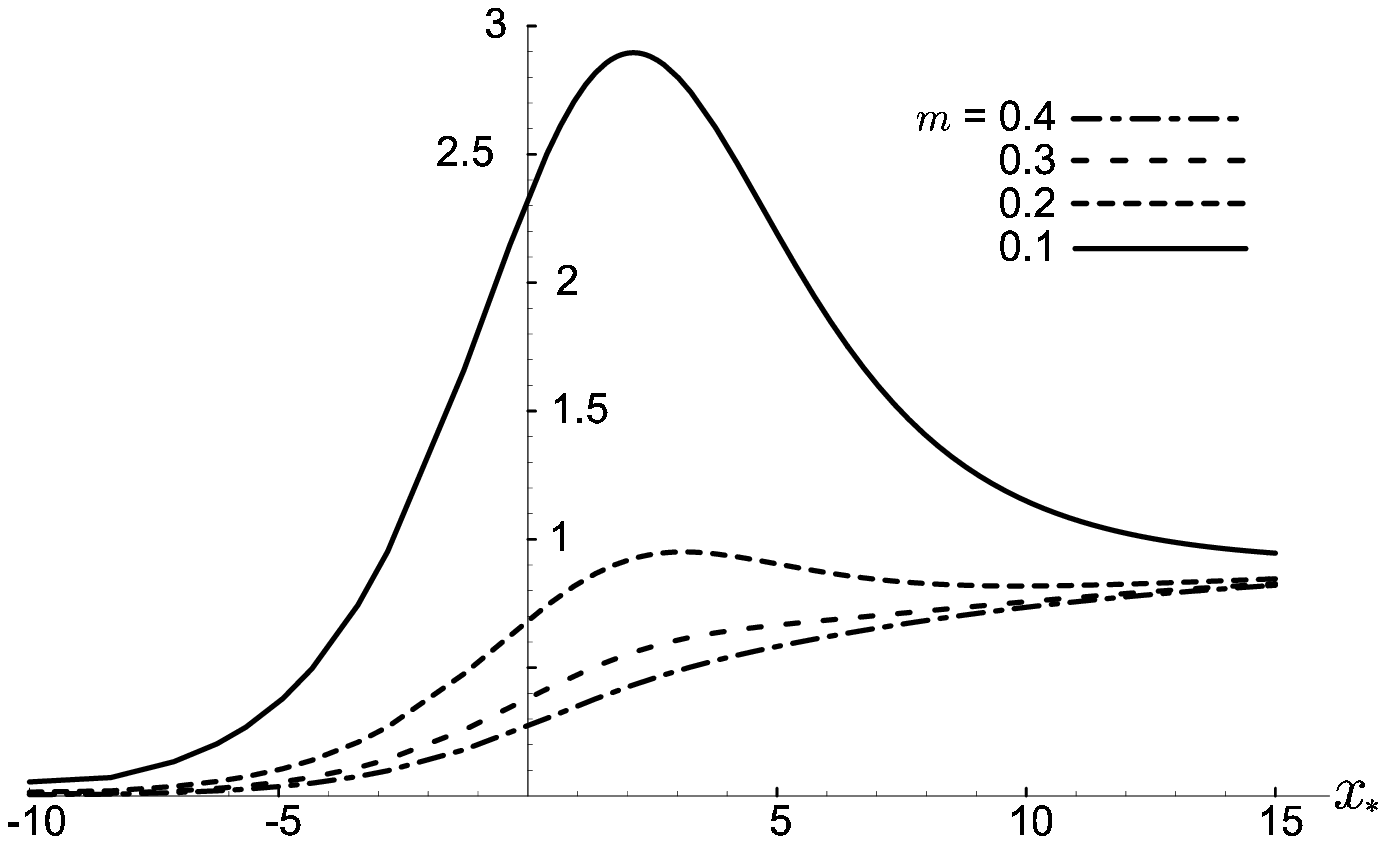}
\caption{\label{caption:potb}
Profiles of the potential (\ref{eq:bhpot}) in a Schwarzschild black hole case
($M=1$).
It is normalised by $m^2$.}
\end{center}
\end{figure}

Let us consider the following model:
\numparts\label{eq:anapot}
\begin{eqnarray}
&&\left[\frac{d^2}{dx_*^2}+\nu^2-V(x_*)\right]Z(x_*)=0 \label{eq:heq} \\
&&{~~~}V=\frac{1}{4M^2}u(1-u)+m^2u \label{eq:hppot} \\
&&{~~~}u=\frac{1}{2}(1+\tanh \frac{x_*}{2L}) \label{eq:xu}.
\end{eqnarray}
\endnumparts
Here, the parameters in the above equations cannot have any physical meanings,
but we can consider some correspondences to the black hole parameters.
$M$ and $m$ correspond to the black hole mass and
field mass, respectively.
$L$ determines the width of the peak and this is dependent
on $M$ in the black hole case.
The coordinate variable $x_*$ corresponds to the so-called
``tortoise'' coordinates.
The potential of this model \eref{eq:hppot} is shown
in \fref{caption:poth}. For comparison with a real black hole one,
we also show the potential for a Schwarzschild black hole.
Because both potentials are very similar, we can also expect that
the behaviours of QNMs will resemble in both cases. 

This model can be solved analytically as follows.
First, by changing the coordinate $x_*$ to $u$,
\eref{eq:heq} can be rewritten into the following form,
\begin{equation}\label{eq:ueq0}
\left[u(1-u)\frac{d^2}{du^2}+(1-2u)\frac{d}{du}+
\frac{L^2\nu^2}{u(1-u)}-\frac{L^2}{4M^2}-\frac{m^2L^2}{1-u}
\right]Z(u)=0.
\end{equation}
Note that the infinity of $x_*=\pm\infty$ corresponds to
$u=1~(+\infty)$ and $0~({-\infty})$.
These two points $u=0$ and $1$ are the regular singularities of
the differential equation \eref{eq:ueq0}.
Then, the behaviour of $Z(u)$ around these two singularities is given by
\begin{equation}
Z(u)\sim u^{a}(1-u)^{b},
\end{equation}
where $a^2=-L^2\nu^2$, $b^2=-L^2(\nu^2-m^2)$.
Therefore, by setting 
\begin{equation}\label{eq:ZZZ}
Z(u)=u^{a}(1-u)^{b}g(u)
\end{equation}
and substituting \eref{eq:ZZZ} into \eref{eq:ueq0},
we obtain a differential equation for $g(u)$ as follows,
\begin{eqnarray}
&&\Biggl[u(1-u)\frac{d^2}{du^2}+\left\{(1+2a)-2(1+a+b)u\right\}\frac{d}{du}
\nonumber \\
&&{~~~~~~~~~~~~}+\left\{-a-b-2ab-L^2m^2-\frac{L^2}{4M^2}+
2L^2\nu^2\right\}\Biggr]g(u)=0.
\end{eqnarray}
This is exactly the hyper-geometric equation and the solution of this
equation can be given by the hyper-geometric function
$F(\alpha,\beta,\gamma;u)$.
In consequence, we find the solution of \eref{eq:ueq0} as follows,
\numparts
\begin{eqnarray}
Z(u)&=&u^a(1-u)^bF(\alpha,\beta,\gamma;u), \\
\alpha,\beta&=&\frac12+a+b\pm\sqrt{1-\left(\frac{L}{M}\right)^2}, \\
\gamma&=&1+2a.
\end{eqnarray}
\endnumparts

Now, let us give our attention to the following relation\cite{SpeFunc},
\begin{eqnarray}\label{eq:trans}
&&u^a(1-u)^bF(\alpha,\beta,\gamma;u) \nonumber \\
&&{}=\frac{\Gamma(\gamma-\alpha-\beta)\Gamma(\gamma)}
{\Gamma(\gamma-\alpha)\Gamma(\gamma-\beta)}u^a(1-u)^b
F(\alpha,\beta,1+\alpha+\beta-\gamma;1-u) \nonumber \\
&&{}+\frac{\Gamma(\gamma)\Gamma(\alpha+\beta-\gamma)}
{\Gamma(\alpha)\Gamma(\beta)}u^a(1-u)^{-b}
F(\gamma-\alpha,\gamma-\beta,1-\alpha-\beta+\gamma;1-u).
\end{eqnarray}
According to \eref{eq:trans},
we can analytically determine QNM frequencies
from the behaviour of the solution around $u=0$ and $1$,
because QNM is defined by the boundary conditions at $x_*=\pm\infty$,
{\it i.e.} $u=0$, $1$.
The boundary condition is that: QNMs are
\begin{enumerate}
\item ingoing waves at $x_*\sim-\infty$ ($u=0$),
\item outgoing waves at $x_*\sim+\infty$ ($u=1$), and
\item damping modes
\footnote{We suppose the time dependency is $\exp(-i\nu t)$}.
\end{enumerate}
From the first condition at $u=0$, we have to choose the sign of $a$ as,
\begin{equation}
a=-iL\nu.
\end{equation}
The second condition at $u=1$ means
\begin{equation}\label{eq:abgamma}
\frac{\Gamma(\gamma)\Gamma(\alpha+\beta-\gamma)}
{\Gamma(\alpha)\Gamma(\beta)}=0,
\end{equation}
because $(1-u)^{-b}$ represents the incoming waves at $u=1$ when
we set $b=+iL\sqrt{\nu^2-m^2}$
\footnote{The branch of $\sqrt{z}$ is $-\pi/2<\arg\sqrt{z}\le\pi/2$.}.
Therefore, QNM frequencies can be determined by,
\begin{equation}
\alpha,~\beta~=-n~(n=0,1,\cdots),
\end{equation}
according to \eref{eq:abgamma}.
By solving this equation for $\nu$, we obtain
\begin{eqnarray}\label{eq:nu}
\nu&=&{}-i\frac{(2n+1)\left(4n^2+4n+\mu^2-4\mu^2\epsilon^2\right)}
{4L\left(4n^2+4n+\mu^2\right)} \nonumber \\
&&{~~~~~}\mp\frac{\left(4n^2+4n+\mu^2+4\mu^2\epsilon^2\right)}
{4L\left(4n^2+4n+\mu^2\right)}\left(i\sqrt{1-\mu^2}\right),
\end{eqnarray}
where we set $L=\mu M$ and $\epsilon=mM$.
When $\mu\leq1$, $\nu$ is pure imaginary.
However, $\nu$ becomes a complex number if $\mu>1$.
In this case, the first term in the right hand side of \eref{eq:nu}
represents the damping part and the second term does
the oscillation part.
Consequently, the damping coefficient of QNMs approaches to zero
as a quadratic function of $\epsilon$.
However, $\nu$ never becomes a real number because
the branch of square root changes discontinuously 
when QNM frequencies get across the real axis.
According to this result, we can conclude that there exist
QNMs with arbitrarily long decaying time when $m$ changes appropriately.
In other words, we find an interesting fact that
there is a kind of resonance mode in a limiting situation
and QNMs can disappear when $m$ exceeds a certain value.

This disappearance is due to
the relation of the height of the peak of
the potential with the field mass.
Therefore, we can expect that
the same phenomenon will occur in the black hole case.
Thus, we will investigate the massive QNMs on the black hole spacetime
in the next section.

\section{Massive QNM on black hole spacetime}\label{sec:bh}
In this section, we consider QNMs of a massive scalar field on
a spherical black hole spacetime. In the black hole case,
we must tackle the problem in a numerical way
because it is difficult to study it fully analytically.

The metric of a spherical symmetric spacetime is given by
\begin{equation}
ds^2=-f(r)dt^2+f^{-1}(r)dr^2+r^2d\Omega^2.
\end{equation}
In this article, we consider the Reisner-Nordstr\"om black hole
as the back ground spacetime, so the metric function $f(r)$ is given by:
\begin{equation}
f(x)=1-\frac{2M}{r}+\frac{Q^2}{r^2}=1-\frac{2}{x}+\frac{q^2}{x^2},
\end{equation}
where $M$ is the black hole mass and $Q$ is the black hole charge.
We also introduce the normalised variables $x$ and $q$ as below,
\numparts
\begin{eqnarray}
x&=&\frac{r}{M}, \\
q&=&\frac{Q}{M},
\end{eqnarray}
\endnumparts
respectively.

In this spacetime, the radial equation for a scalar field
in term of the normalised variables is given by
\begin{eqnarray}\label{eq:Z}
\left[\frac{d^2}{dx^2}+\frac{\frac{df}{dx}}{f}\frac{d}{dx}+
\left(\frac{\nu^2}{f^2}-\frac{l(l+1)}{x^2f}-
\frac{m^2}{f}-\frac{\frac{df}{dx}}{xf}\right)\right]Z(x)=0,
\end{eqnarray}
where $\nu$ and $\epsilon$ are normalised frequency and mass of
the scalar field and they are related to 
original frequency $\omega$ and mass $m$
\footnote{The original field equation is $(\Box-m^2)\phi=0$, and
$\omega$ is introduced by variable separation,
$\phi\sim e^{-i\omega t}Z(r)Y_{lm}(\theta,\varphi)$.} by
\numparts
\begin{eqnarray}
\nu&=&M\omega, \\
\epsilon&=&mM,
\end{eqnarray}
\endnumparts
respectively.

\begin{figure}
\begin{center}
\includegraphics[height=60mm]{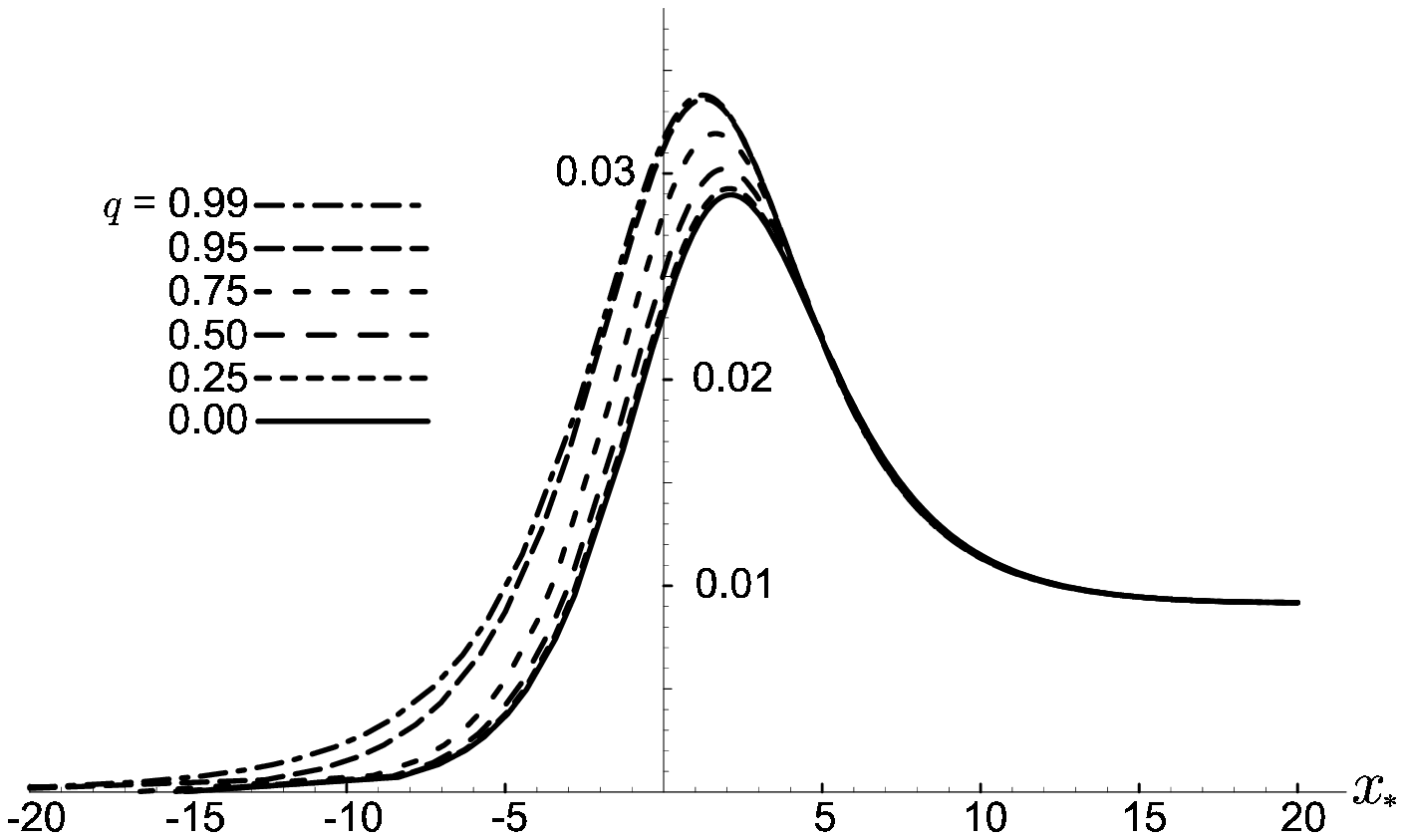}
\caption{\label{caption:potm01}Potential in case of $\epsilon=0.1$}
\vspace{4mm}
\includegraphics[height=60mm]{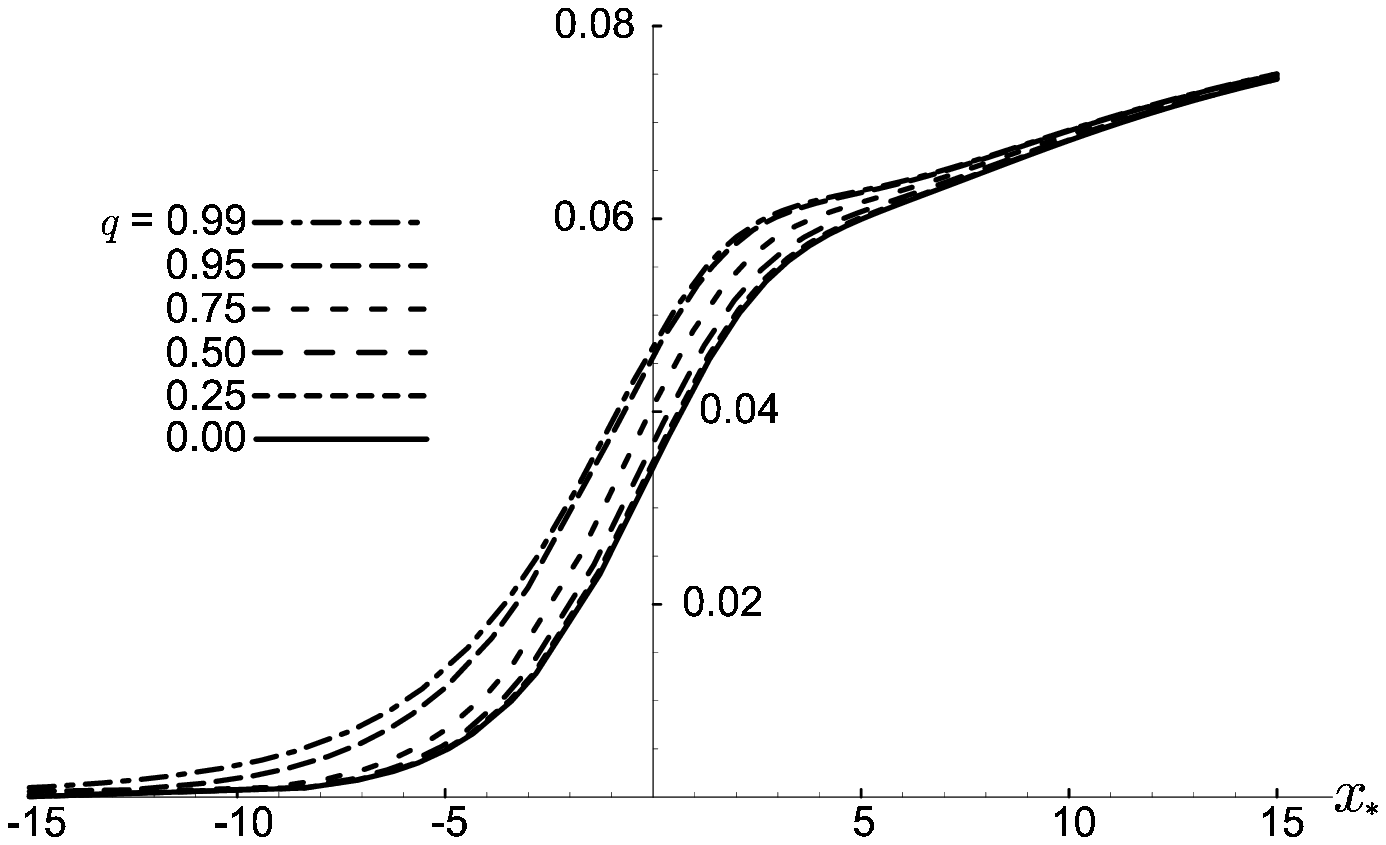}
\caption{\label{caption:potm03}Potential in case of $\epsilon=0.3$}
\end{center}
\end{figure}
In order to find the potential for the scalar field,
let us introduce a so-called tortoise coordinate $x_*$ by
\begin{equation}
\frac{dx_*}{dx}=\frac1f \label{eq:xstar},
\end{equation}
and we obtain the following equation,
\begin{equation}\label{eq:RNscalar}
\left[\frac{d^2}{dx_*^2}+\nu^2-V(x_*)\right]Z=0,
\end{equation}
where the potential $V(x_*)$ is given by
\begin{equation}\label{eq:bhpot}
V(x_*)=\left(\frac{l(l+1)}{x^2}+\epsilon^2+\frac{\frac{df}{dx}}{x}\right)
\times f(x).
\end{equation}
Note that $x$ in the right hand side of \eref{eq:bhpot}
is a function of $x_*$ by \eref{eq:xstar}.
The profiles of this potential in some cases are shown in 
\fref{caption:potm01} and \fref{caption:potm03}.
Figure \ref{caption:potb} in \sref{sec:am} also shows
the potential for comparison with the potential of the analytic model.

In order to calculate the QNM frequencies,
we solve \eref{eq:Z} numerically.
We consider the case of $|q|<1$ in this article,
because the mathematical structure of \eref{eq:Z} with $q=1$
is apparently different from the case of $|q|<1$.
The reason of excluding the case of $|q|>1$ is that
there is no horizon in this case and then
the central singularity is naked.
In this case, QNMs cannot be defined.
Although the Schwarzschild case, $q=0$, seems to be different case,
we verify that there is no difference between the two cases in our results:
(i) $q=0$ and (ii) $q\to0$.

\begin{figure}
\begin{center}
\includegraphics[height=60mm]{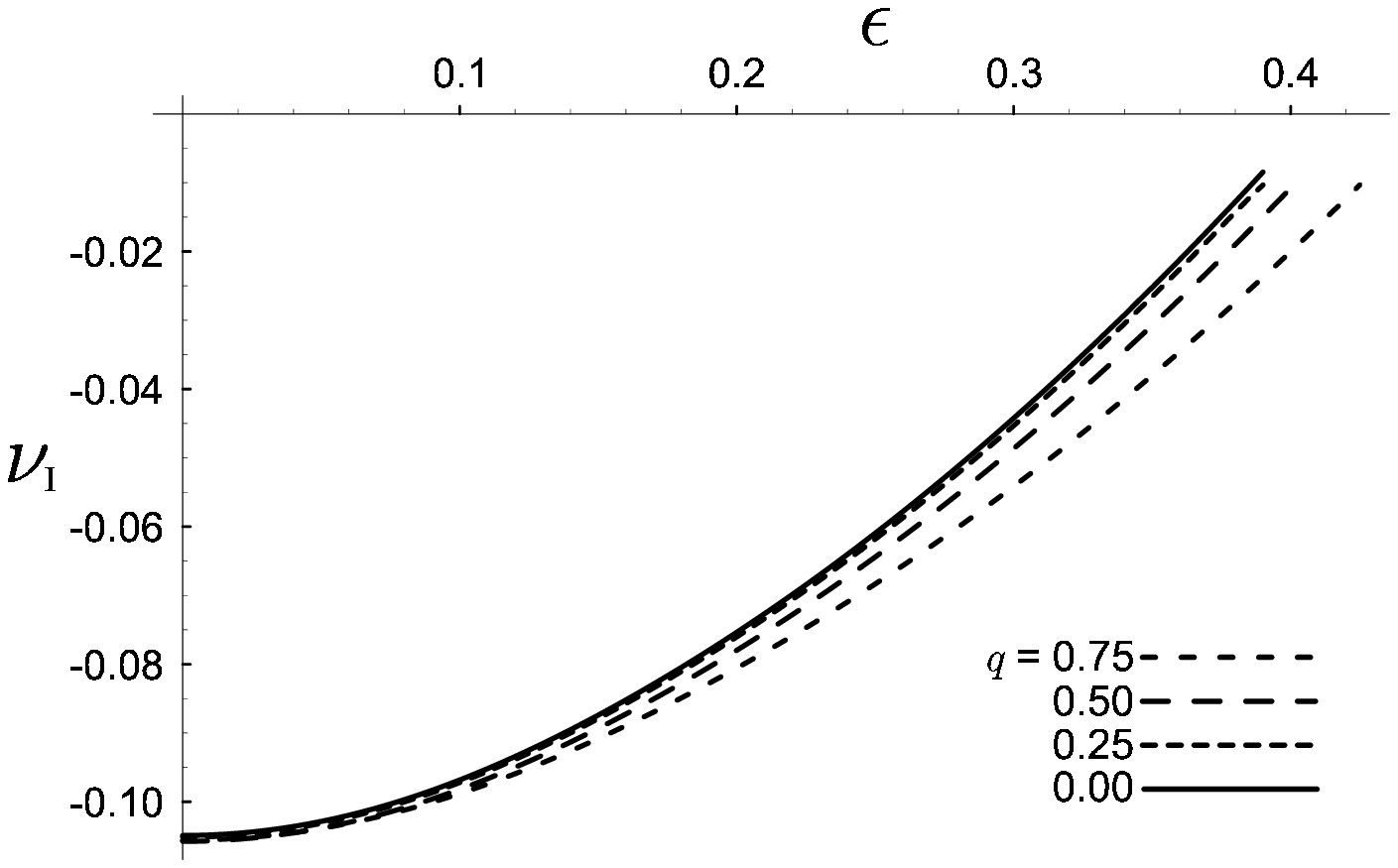}
\caption{\label{caption:reqnm1}Dependence of
${\rm Im}(\nu)$ to the field mass $\epsilon$
for the cases of $q=0.0$, $0.25$, $0.50$ and $0.75$.}
\vspace{4mm}
\includegraphics[height=60mm]{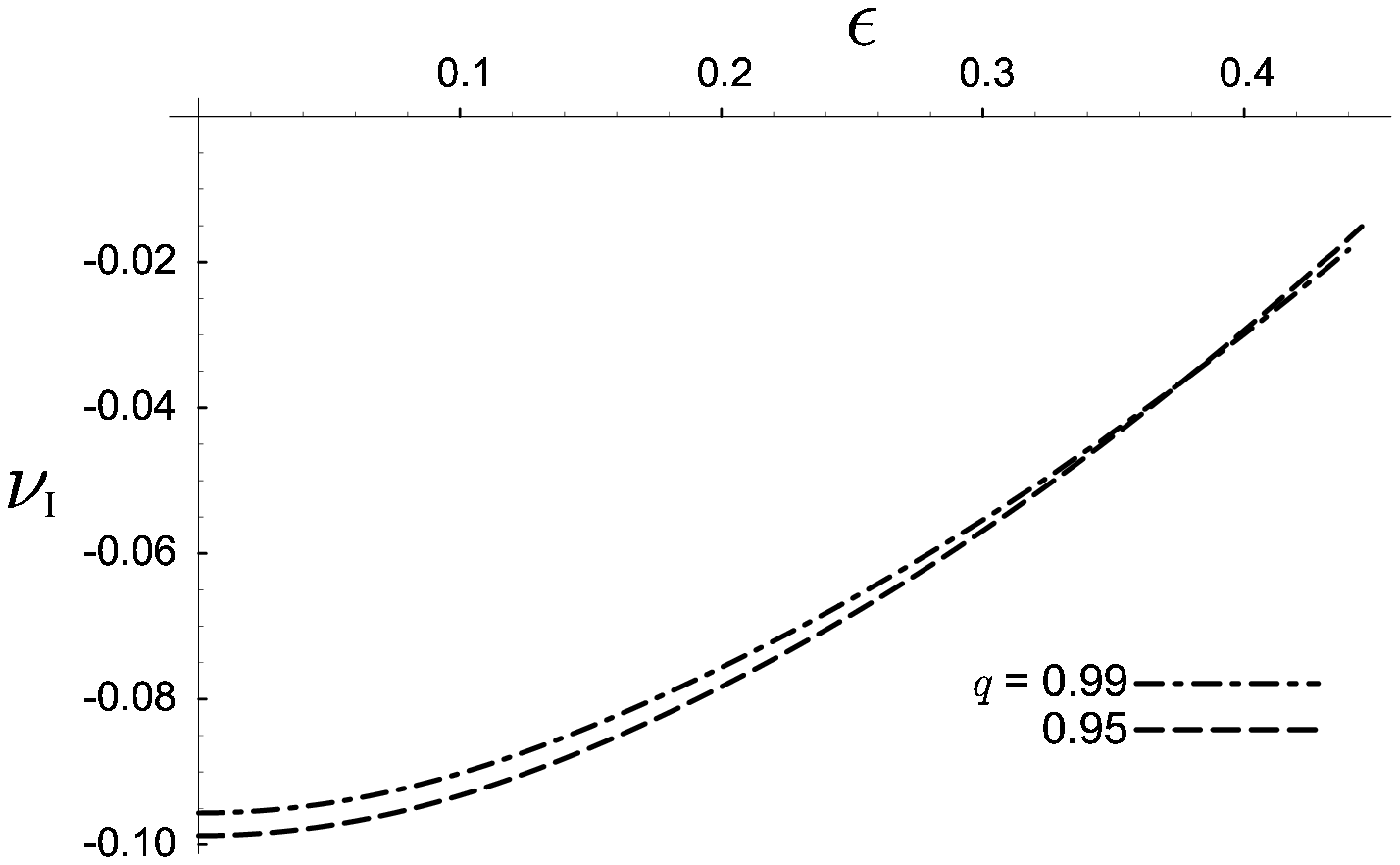}
\caption{\label{caption:reqnm2}Dependence of
${\rm Im}(\nu)$ to the field mass $\epsilon$
for the cases of $q=0.95$ and $0.99$.}
\end{center}
\end{figure}

\begin{figure}
\begin{center}
\includegraphics[height=60mm]{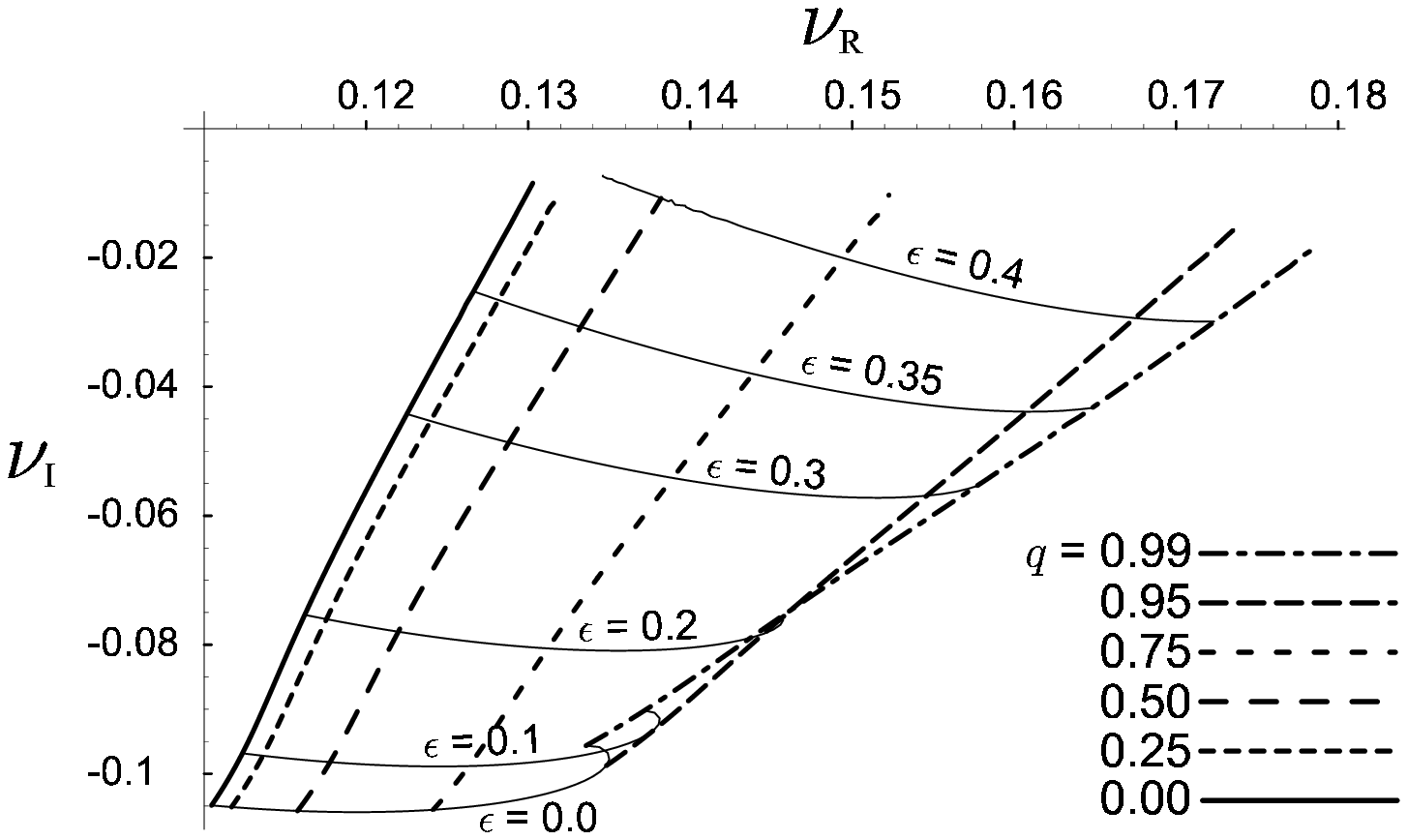}
\caption{\label{caption:reqnm3}Trajectory of QNM frequencies
in $\nu$-plane.}
\vspace{4mm}
\includegraphics[height=60mm]{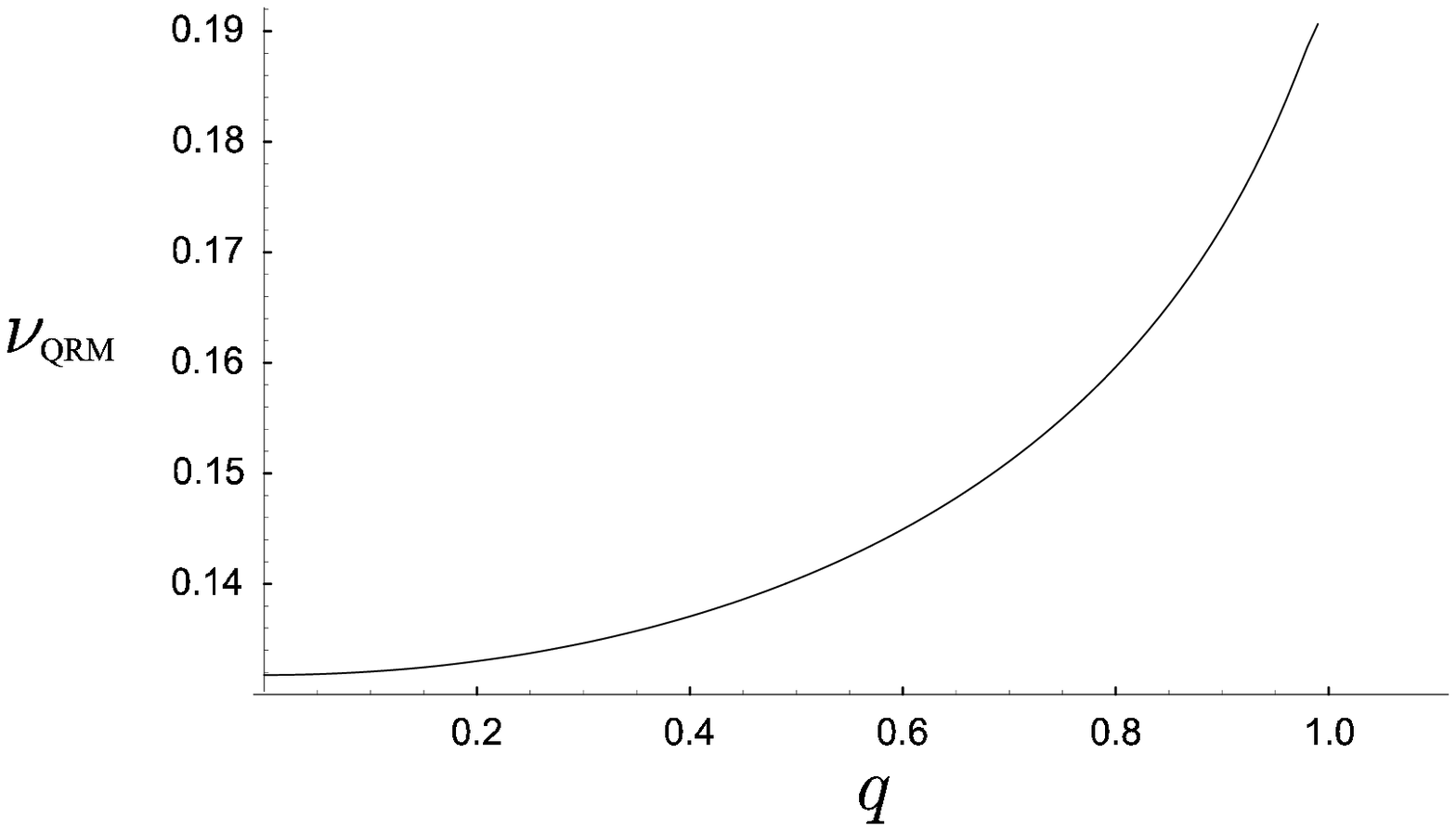}
\caption{\label{caption:qrm1}Dependence of QRM frequencies on
the charge of the black hole.}
\end{center}
\end{figure}

The numerical method which we use in calculating QNM frequencies
is the continued fraction method developed by Leaver\cite{Leaver}.
In \ref{sec:Leaver}, we will briefly review this method for our case.
In this article, we examine only $l=0$ and calculate
QNM frequencies from $\epsilon=0$ to around $0.4$
in several cases of $q=0~({\rm Schwarzschild~case})$,
$0.25$, $0.5$, $0.75$, $0.95$ and $0.99$.
The results are shown in \fref{caption:reqnm1}, \fref{caption:reqnm2},
\fref{caption:reqnm3} and \fref{caption:qrm1}.
In these figures, $\nu_{\rm I}$ and $\nu_{\rm R}$
mean the imaginary and real part of the QNM frequencies, respectively.

Figures \ref{caption:reqnm1} and \ref{caption:reqnm2} show
the dependence of the QNM frequencies on the mass of the scalar field.
It is obvious from the figures that
the imaginary part of the QNMs on the black hole spacetime approaches to $0$
as well as the analytical model in the last section.
This tendency is more clear in figure \ref{caption:reqnm3}.
Figure \ref{caption:reqnm3} shows the trajectories of the QNM frequencies on
the $\nu$-plane when the field mass $\epsilon$ changes
while keeping $q$ constant.
From this figure, we can find that
the trajectories of the QNM frequencies are almost linear,
and we can guess the frequencies when QNMs disappear.
We call the modes with these {\it real} frequencies as
{\it quasi-resonance modes (QRM)}.
According to the linearity of the QNM trajectories,
we can calculate the dependence of QRM frequencies
on the field mass $\epsilon$ with the following fitting function,
\begin{equation}
\nu_{\rm I}=k(\nu_{\rm R}-\nu_{\rm QRM}).
\end{equation}
The behaviour of this $\nu_{\rm QRM}$ to $\epsilon$ is
plotted in \fref{caption:qrm1}.

We will give more detailed discussion on our results in the next section.

\section{Discussion}\label{sec:ds}
In this article, we have studied massive QNMs for the two cases,
the analytically solvable model and the black hole.
We can find the same phenomenon, that
the life of QNMs becomes longer as the field mass increases,
in both cases.
From this fact, we can conclude that
there exists a QNM, oscillating without damping for an arbitrary period.
This extreme mode may be regarded as a kind of resonance mode
(after this fact, we named this mode {\it quasi-resonance mode, QRM},
in the last section).
Now, we propose a physical picture of the disappearance of QNM
and the existence of QRMs.

First, let us discuss the case of the analytic model.
We consider only $n=0$ case for simplicity.
By setting $n=0$ in \eref{eq:nu}, we obtain
\begin{eqnarray}
\nu&=&{}-i\frac{1-4\epsilon^2}{4L}\mp\frac{1+4\epsilon^2}{4L}
\left(i\sqrt{1-\mu^2}\right).
\label{eq:nu_n=0}
\end{eqnarray}
What is noticed from \eref{eq:nu_n=0} is that
QNM frequencies are almost determined by the width $L$ of the potential peak
when $\epsilon\sim0$, and they turn out to depend on $\epsilon^2$
as $\epsilon$ increases.
Since the energy of QNMs is proportional to the real part of $\nu$,
the larger the energy of QNMs is,
the gentler the peak of the potential becomes.
Then the peak will vanish at $\epsilon=1/2$.
According to the WKB analysis of QNM by Schutz and Will \cite{WKB},
QNMs can be regarded as the waves trapped by the peak of the potential.
Therefore, in their picture, vanishing of the peak is equivalent to
the potential being unable to trap any waves.
This means that QNMs disappear.

\begin{figure}
\begin{center}
\includegraphics[height=60mm]{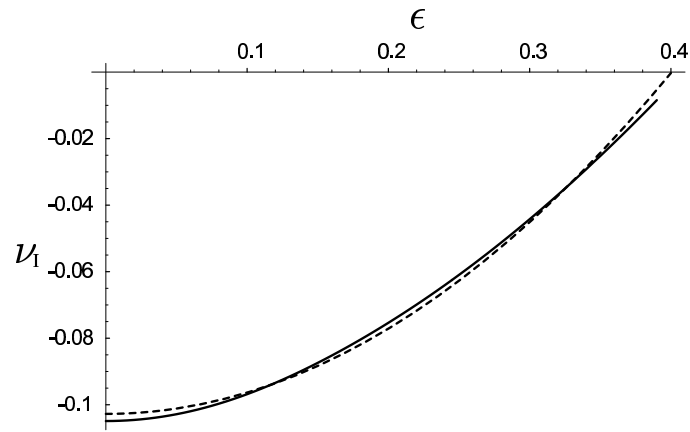}
\caption{\label{caption:fitting}Fitting by a parabola
$\nu_{\rm I}=c_0+c_2 \epsilon^2$.
The solid line shows the QNM trajectory of $q=0$ and the dashing line
shows the fitting curve.}
\end{center}
\end{figure}
Next, let us consider the black hole case in this picture.
In the black hole case, the peak of the potential vanishes at, for example,
$\epsilon=0.25$ for $q=0$ and $\epsilon\sim0.256$ for $q=0.5$.
On the other hand, \fref{caption:reqnm1} and \fref{caption:reqnm2}
indicate disappearance of QNMs approximately
at $\epsilon=0.4\sim0.5$.
Thus, the above picture seems to be realized approximately.
Because the QNM frequencies are {\it complex numbers}, 
it is unreasonable to think that
the disappearance of QNMs perfectly coincides with vanishing
of the peak of the {\it real} potential in general.
Hence, it is natural to consider that the coincidence
in the analytic model is accidental.
However, the behaviours of QNMs to $m$ in both cases
resemble each other very well.
Figure \ref{caption:fitting} shows that
the trajectory of QNMs can be fitted by a parabola very well.
That is, our analytic model imitates the black hole case precisely.
Therefore, since the analytic model predicts the disappearance of QNMs,
it is undeniable that there exist QRMs in the case of black hole
as well as the case of analytic model,
and that QNMs will disappear as the field mass exceed a certain value.

We also examine the effect of the black hole charge.
We can see the behaviours of QNMs and QRMs
when we change the black hole charge, $q$,
in \fref{caption:reqnm3} and \fref{caption:qrm1}.
These figures indicate that QNM and QRM frequencies
tend to get larger when $q$ increases.
This can be explained by the potential profiles.
According to \fref{caption:potm01} and \ref{caption:potm03},
when $q$ increases, the peak of the potential becomes higher and
the value of $\epsilon$ at vanishing point of the peak becomes larger. 
Consequently, QNMs can exist at larger $\epsilon$ and
QRM frequencies get higher.
These facts, too, support our picture of the disappearance of QNMs.

\section{Summary}\label{sec:sum}
Let us summarise our results.
First, we studied massive QNMs
in the analytical model, and derived
QNM frequencies analytically.
By this analysis, we found that QNMs will disappear when
the field mass increases and exceeds a certain value.
Secondly, we examined the dependence of QNM frequencies on
the field mass
by numerical method in the case of black hole.
There, we again found the disappearance of QNMs
as we did in the analytic model case.
We also studied the dependence of QNM frequencies
on the charge of the black hole.
Based on these examinations, we proposed a picture
about the disappearance of QNMs.

\appendix
\section{Continued fraction method}\label{sec:Leaver}
In the appendix, we briefly review the continued fraction method,
which is first developed by Leaver\cite{Leaver}.

\begin{figure}
\begin{center}
\includegraphics[height=60mm]{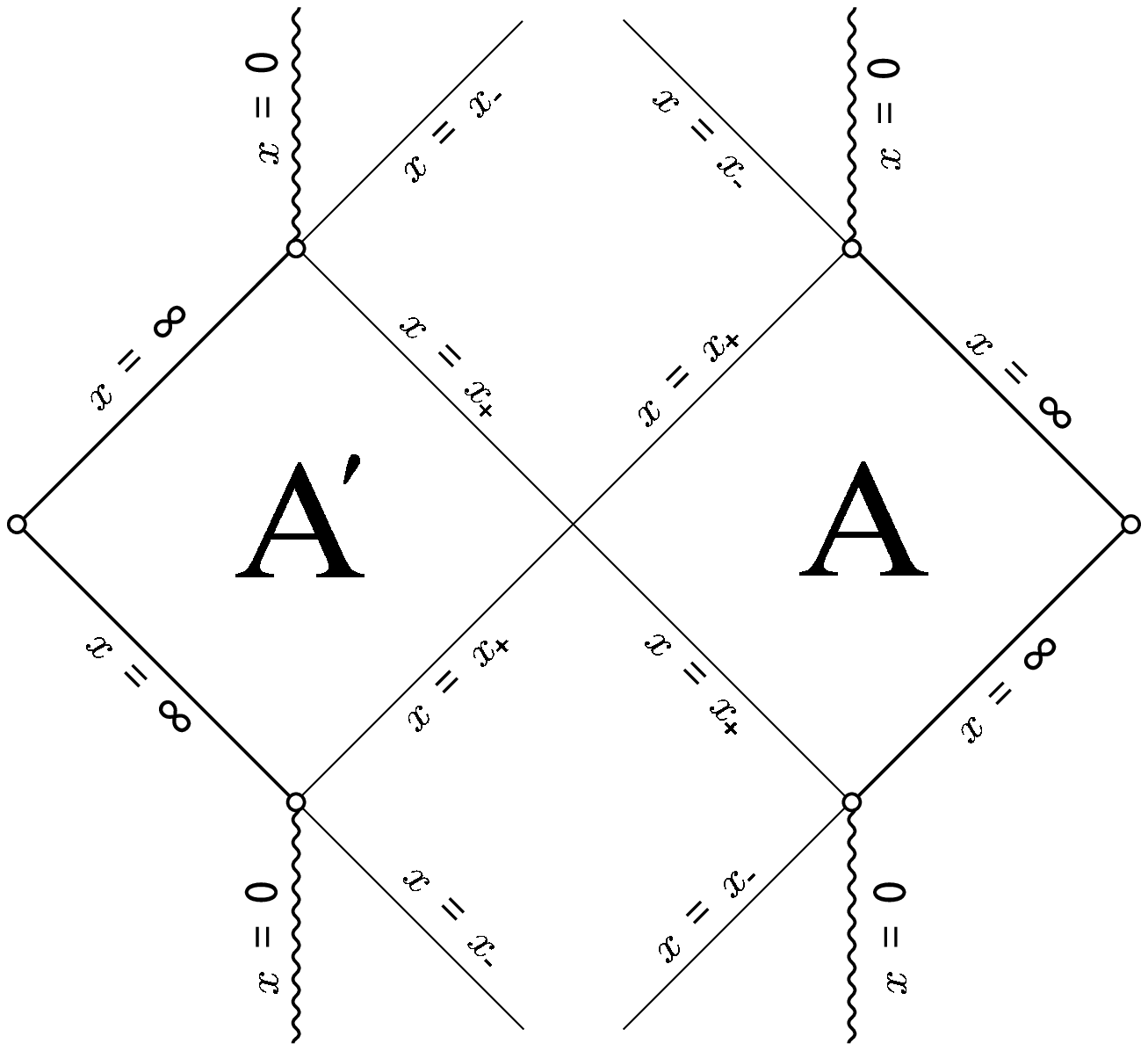}
\caption{\label{caption:rnconf}
Conformal diagram for a Reisner-Nordstr\"om black hole
with $|q|<1$.}
\end{center}
\end{figure}

First, let $x_{\pm}$ be the solution of $f(x)=0$ in \eref{eq:Z}.
Then, the points $x=0,~x_-,~x_+$ are regular singularities
and the point $x=\infty$ is an irregular singularity
of the differential equation \eref{eq:Z}.
Note also that $x=x_-$ corresponds to the inner horizon
of the black hole and $x=x_+$ corresponds to the outer horizon.
QNMs on a Reisner-Nordstr\"om black hole spacetime exist
in regions connected to the spatial infinity of the spacetime.
These regions are shown as A, A' in the conformal diagram
(\fref{caption:rnconf}).
Therefore, the boundary conditions for QNMs on this spacetime
are imposed at $x=x_+$ and at $x=\infty$.

Let us expand the solution of \eref{eq:Z} as follows,
\begin{equation}\label{eq:zexpand}
Z(u)=e^{i\nu x}u^{\rho}(x-x_-)^{b}x
\sum_{n=1}^{\infty}a_nu^n,
\end{equation}
where 
\begin{eqnarray*}
u=\frac{x-x_+}{x-x_-}, \\
\nu^2=\chi^2-\epsilon^2, \\
\rho=-i\chi\frac{x_+{}^2}{x_+-x_-}, \\
b=-1+\frac{i}{2\nu}(\nu^2+\chi^2)(x_++x_-).
\end{eqnarray*}
Substituting \eref{eq:zexpand} into \eref{eq:Z},
we obtain the following recursion relation,
\begin{eqnarray}
A_1a_1+B_1a_0=0, && \\
A_{n}a_{n}+B_{n}a_{n-1}+C_{n}a_{n-2}=0, &&{\rm ~~~for~}n=2,3,\cdots,
\end{eqnarray}
where
\begin{eqnarray}
A_{n}&=&n^2(x_{-}-x_{+})+2inx_{+}{}^2\chi, \\
B_{n}&=&{}-2n^2(x_{-}-x_{+}) \nonumber \\
    &&{}+n\times\Biggl\{
        i(x_{-}-x_{+})\Bigl[\frac{\chi^2}{\nu}(x_{-}+x_{+})
            -\nu(x_{-}-3x_{+})
        \Bigr] \nonumber \\
    &&{~~~~~~~~~~~}+2(x_{-}-x_{+}-2ix_{+}{}^2\chi)
    \Biggr\} \nonumber \\
    &&{}+\Biggl\{
    -\bigl[l(l+1)+1\bigr](x_{-}-x_{+}) \nonumber \\
    &&{~~~~~~}+\bigl[\nu^2(x_{-}-x_{+})+2i\chi-(x_{-}+3x_{+})\chi^2
    \bigr]x_{+}{}^2 \nonumber \\
    &&{~~~~~~}-\frac{i}{2}(x_{-}-x_{+}-2ix_{+}{}^2\chi)
    \Bigl[\frac{\chi^2}{\nu}(x_{-}+x_{+})-\nu(x_{-}-3x_{+})\Bigr]
    \Biggr\}, \\
C_{n}&=&n^2(x_{-}-x_{+}) \nonumber \\
    &&{}+n\times\Bigl[-i\Bigl(\nu+\frac{\chi^2}{\nu}\Bigr)
    (x_{-}{}^2-x_{+}{}^2)
    -2(x_{-}-x_{+}-ix_{+}{}^2\chi)\Bigr]
    \nonumber \\
    &&{}+\Biggl\{
    -\frac{1}{4}\Bigl(\nu^2+\frac{\chi^4}{\nu^2}\Bigr)
    (x_{-}{}^2-x_{+}{}^2)(x_{-}+x_{+}) \nonumber \\
    &&{~~~~~~}+i\Bigl(\nu+\frac{\chi^2}{\nu}\Bigr)(x_{-}+x_{+})
    (x_{-}-x_{+}-ix_{+}{}^2\chi) \nonumber \\
    &&{~~~~~~}+\Bigl[(x_{-}-x_{+})-2ix_{+}{}^2\chi
    +\frac{1}{2}(x_{-}+x_{+})(x_{-}{}^2+3x_{+}{}^2)\chi^2\Bigr]  
    \Biggr\}.
\end{eqnarray}
Using these coefficients $A_{n}$, $B_{n}$ and $C_{n}$,
QNM frequencies are given by the vanishing point of
the following continued fraction equation,
\begin{equation}
0=B_1-\frac{A_1C_2}{B_2-}\frac{A_2C_3}{B_3-}\cdots
\frac{A_{n-1}C_n}{B_n-}\cdots.
\end{equation}
For practical calculation,
we must choose $n$ for the needed accuracy and
solve the algebraic equation of finite but very large degrees.

\end{document}